# FDH knockout and *Ts*FDH transformation led to enhanced growth rate of *Escherichia coli*

Roya Razavipour[1†], Saman Hosseini Ashtiani[2†*], Abbas Akhavan Sepahy[3], Mohammad Hossein Modarressi[4], Arne Elofsson[2], Bijan Bambai[5*]

1. Department of Biology, Science and Research Branch IAU, 1477893855, Tehran, Iran, roya.razavipoor@gmail.com
2. Department of Biochemistry and Biophysics and Science for Life Laboratory, Stockholm University, 106 91 Stockholm, Sweden, ORCID: 0000-0002-7115-9751, arne@bioinfo.se
3  Department of Microbiology, Faculty of Science, North Branch IAU, 1651153311, Tehran, Iran, akhavansepahi@gmail.com
4. Department of Genetics, Medical School, Tehran University of Medical Sciences, 1416753955 Tehran, Iran, modaresi@tums.ac.ir
5. Department of Systems biotechnology, Faculty of Industrial and Environmental Biotechnology, National Institute for Genetic Engineering and Biotechnology (NIGEB), 1497716316 Tehran, Iran, bambai@nigeb.ac.ir

[*]Corresponding Authors: Bijan Bambai, bambai@nigeb.ac.ir Tel: +982144787320 and Saman Hosseini Ashtiani, saman.hosseini-ashtiani@dbb.su.se, ORCID ID: 0000-0003-2381-3410; Tel.: +46-762623644
† Have equally contributed to the manuscript as the first author

# Abstract

Background:
Increased Atmospheric $CO_2$ to over 400 ppm has prompted global climate irregularities. Reducing the released $CO_2$ from biotechnological processes could remediate these phenomena. In this study, we sought to reduce the released $CO_2$ into the atmosphere from bacterial growth by reducing formic acid conversion into $CO_2$. Since *E. coli* is the biotechnological workhorse and its higher growth rate is desirable, another goal was to monitor the bacterial biomass after the metabolic engineering.

Results:
The biochemical conversion of formic acid to $CO_2$ is a key reaction. Therefore, we compared the growth of control strains K12 and BL21, alongside two strains (in which two different genes coding two formate dehydrogenase (FDH) subunits were deleted) in complex and simple media. Our observations demonstrated that the knockout bacteria significantly grew more efficiently than the controls in both media. *Ts*FDH, an FDH with moderately more catalytic efficiency, in contrast to other known FDHs for converting $CO_2$ to formate, increased the growth of both knockouts compared with the controls and the knockouts without *Ts*FDH. This difference was more accentuated in M9+Glycerol. Through a transcriptomics-level *in silico* analysis of the knockout genes, RNA-seq-based correlation outcome revealed that the genes negatively correlated with the target genes (knockout genes) belong to tRNA-related pathways.

Conclusion:
Observing higher cell biomass for the knockout and transformed strains at equal concentrations of carbon source in both media indicates possible underlying mechanisms leading to reduced carbon leakage and increased carbon assimilation, which need more detailed investigations. These results may also provide a phenotypic-level clue for the inconsistency of predictions in previous metabolic models that declared glycerol as a suitable carbon source for the growth of *E. coli* but failed to achieve it in practice. Gene expression correlations and pathway analysis outcomes suggested possible over-expression of the genes involved in tRNA processing and charging pathways.

# Introduction

$CO_2$ is easily formed by the oxidation of organic molecules during respiration in living organisms or combustion in regular mechanical engines. This molecule is thermodynamically stable with a low chemical activity. Today, the atmospheric $CO_2$ concentration is approaching alarming levels (from 300 ppm to 417 *ppm* in about 50 years). This has led to elevated frequencies of extreme climate conditions like drought, flooding, wild fire and tropical storms in different regions of the world (1). The development of innovative methods for reducing the released $CO_2$ into the atmosphere and assimilating $CO_2$ into organic matter is in demand more than ever (2).

Respiration is a more economical process for extracting chemical energy from organic matter compared with anaerobic fermentation. There is still an undesired side effect of respiration, i.e., the release of $CO_2$. Engineering bacterial strains involved in biotechnological processes with the aim of reducing carbon dioxide release into the atmosphere or even fixing the atmospheric $CO_2$ into biomass has environmental and economic advantages. There have been a number of efforts to reduce the carbon dioxide release during biomass production by metabolic engineering (3). The central pathways and cycles of metabolism are the first targets for manipulating enzymes responsible for critical biochemical reactions or regulatory proteins controlling the expression of certain enzymes to reduce $CO_2$ release (4).

One of the interesting candidates for reducing the amount of released $CO_2$ is formate dehydrogenase (FDH). Theoretically, FDHs are enzymes capable of reversible conversion of $CO_2$ to formate, which is the simplest organic acid (5). However, the major drawback of the biotechnological application of FDHs is the fact that the majority of these enzymes favor the oxidation of formate to produce $CO_2$ under physiological conditions (6). There are three known FDHs in *E. coli* genome, namely, FdhH, FdhN, and FdhO. The newly identified pressure induced FDH (FHL) is another identified FDH in *E. coli.* In search of an "ideal" FDH we chose to express FDH from *Thiobacillus sp*. KNK65MA (*Ts*FDH) as an enzyme with high catalytic efficiency (7). The crystal structure of active enzyme (PDB: 3WR5) shows a homo-tetramer of 406 amino acid-long polypeptide. There are 5 extra residues at N-terminal of the recombinant protein, compared with the sequence in UniProt (8) (accession code: Q76EB7).

In this study we sought to monitor the growth of *E. coli* fdhD and fdhF knockout strains (JW3866 and JW4040, respectively) with and without *Ts*FDH, as well as control strains, i.e., K12 and BL21. For consistency's sake, we will refer to knockout strains JW3866 and JW4040 as Δ*fdhD* and Δ*fdhF*, respectively. Both knockout strains demonstrated growth advantage in LB as well as M9 + glycerol media compared with the control *E. coli* strains with wild type FDH. Our observations demonstrate a clear growth rate advantage in knockouts expressing the recombinant enzyme (*Ts*FDH) compared with knockouts and controls without *Ts*FDH, particularly in M9+Glycerol medium the difference was higher. In order to perform an *in silico* study related to our observations, we opted for examining the transcriptomic-level correlations between the target (knockout) genes and the rest of the genes in *E. coli*. The correlation analysis based on an independent *E. coli.* RNA-seq gene expression profile data set followed by pathway analysis disclosed that all the genes significantly anti-correlated with both target genes belong to tRNA charging or tRNA processing pathways.

# Materials and Methods

## *Escherichia coli* Strains, Plasmids and Media

All *E. coli* strains and media used in this study are presented in Table 1. *Escherichia coli* BL21(DE3) was used for the expression of the recombinant FDH from *Thiobacillus sp.* KNK65MA (*Ts*FDH). Two FDH knockout strains, Δ*fdhD* and Δ*fdhF* were purchased from Keio Collection.

*Table 1. Bacterial strains and media. All bacterial strains and plasmids used in this study and their characteristics as well as the sources they were obtained.*

| Strains and plasmids | Related characteristics | Source |
|---|---|---|
| **Strains** | | |
| E. coli K12 | Wild type | NIGEB stocks |
| *E. coli* BL21(DE3) | [*lon*] *omp*T *gal* (λ DE3) [*dcm*] Δ*hsd*S | Invitrogen |
| *E. coli* JW3866 | K12 Δ*fdhD* | Dharmacon |
| *E. coli* JW4040 | K12 Δ*fdhF* | Dharmacon |
| **Plasmids** | | |
| pET-21α | Ap$^R$, *T7* promoter, lac operator | Novagen |
| pET-21α-TsFDH | pET-21α, containing TsFDH gene from *Thiobacilusis sp* KNK65MA | This research |

M9 medium with glycerol as carbon source and LB medium as a complex medium all containing 30 µg kanamycin were used for measuring bacterial growth. For BL21 with pET+TsFDH, the same media with ampicillin were used. In all samples containing pET+TsFDH, IPTG (0.5 mM final concentration) was added to the medium. The metabolic reactions consuming or producing formate (map01200 and C00058) were obtained from KEGG (9) (https://www.genome.jp/pathway/map01200+C00058). Using the results from KEGG pathway search for all the carbon fixation reactions, the contributing FDHs were identified. The kinetic parameters including $K_{cat}$ and $K_m$ of FDHs (EC: 1.17.1.9) for formate formation were obtained from Brenda enzyme data bank (10) and the published articles were reviewed and compared in different bacteria (Table S1). This approach revealed some interesting FDHs with relatively better kinetic parameters.

Although, the results obtained by *Ts*FDH might be interesting, we assume there are still some FDHs that deserve attention for replacing the indigenous FDHs of *E. coli* to improve the growth efficiency. Our mentioned assumption is based on the ambiguity of assay conditions for some of the reported FDHs and the lack of a gold standard for the kinetics comparisons. Scanning the kinetic parameters for a desired FDH suggested *Thiobacillus sp*. KNK65MA FDH (7).

Amino acid and nucleotide sequences of *Thiobacilusis sp* KNK65MA formate dehydrogenase were obtained from UniProt (accession # Q76EB7). cDNA of TsFDH was synthetized in pET21a by ZistEghtesadMad based on reference sequence (Q76EB7). Two knockout strains of K12 *Escherichia coli*, Δ*fdhD* and Δ*fdhF*, with the deletion of fdhD and fdhF genes, respectively, were purchased from Dharmacon. The stocks of the Knockout *E. coli* strains were cultured on LB broth and M9+Glycerol media followed by incubation at 37°C for 24 hours (11).

Strains K12 and BL21 were used as control to compare the growth rates. All strains were cultured at the same time under the same conditions on LB broth media at 37 °C and 200 rpm. Competent cells of the BL21, *E. coli* Δ*fdhF*, and *E. coli* Δ*fdhD* were prepared as previously mentioned (11). pET21, a plasmid containing a fusion gene to express format dehydrogenase of *Thiobacillus sp*. KNK65MA (pET+TsFDH), was transformed in competent BL21cells, *E. coli* Δ*fdhD* and *E. coli* Δ*fdhF* on LB Agar with Amp (100 mg / ml) followed by incubation overnight at 37°C. The colonies containing plasmid were selected and cultured on a 10 ml LB broth with Amp as a primary culture and were incubated at 37°C, 200 rpm for 24 hours. Then the culture was carried out in 200 ml of the LB broth containing Amp (100 mg / ml) and they were incubated at 37°C and at 200 rpm for 24 hours. Also, the bacteria K12, BL21, *E. coli* Δ*fdhD* and *E. coli* Δ*fdhF* lacking the plasmid were simultaneously cultivated and incubated on LB broth and M9-Glycerol + 50 µg/ml Kanamycin under identical conditions with plasmid-containing strains.

## Media and culture conditions

M9 medium + glycerol containing 30 µg kanamycin was used for measuring bacterial growth with and without pET+TsFDH. The same media with ampicillin were also used. In all BL21 samples containing pET+TsFDH, IPTG (0.5 mM final concentration) was added to the medium. LB media were purchased from Merck.

## Growth measurements

Bacteria were grown in batch cultures at 37°C in shaker incubator in 50 mL flasks. 1000 µL samples were taken in triplicate at indicated time intervals and the absorbance was measured at 600nm. During incubation, plasmid-free bacteria and plasmid-containing ones were sampled at different times, namely 0h, 2h, 4h, 6h, 8h, 10h, 12h and 24hr (Table S2). To determine the growth rates of bacteria at above time intervals optical absorption was measured using a spectrophotometer at 600 nm wavelength,.

## *In silico* analysis

### Data preparation

To perform a transcriptomic-level study related to our observations, we searched for an independent *E. coli* expression dataset which could reflect the maximum possible transcriptional variations so that we would be able to achieve significant correlations between as many genes as possible. Moreover, the number of genes involved in the gene expression profile was important to calculate as many correlations as possible. With this aspiration, we fetched an *E. coli* RNA-seq dataset comprising 152 RNA-seq count samples under 34 different growth conditions (GEO accession GSE94117). These samples were taken from both exponential and stationary phases. One unique aspect of this highly pertinent dataset is the fact that it is sampled under 34 different growth conditions leading to a wider range of differentially expressed genes thanks to different metabolic needs (12). Subsequently, correlation analysis, PCA and pathway analysis were applied to this data set.

### Data preprocessing

Using Python version 3.6.1, 152 samples of RNA-seq count files were merge. The counts were converted into count per million (CPM) and were log2 transformed. The resulting data were z-score transformed per gene across all samples. Quality control was performed as sample-level box plots before and after data preprocessing (Fig. S1 and S2).

### Correlation analysis

The Spearman rank-order correlation coefficient (Spearman's ρ), being a nonparametric measure, examines the monotonic relationship between the ordinal values of the variables. Contrary to the Pearson correlation, the Spearman's rank correlation is not based on the assumption that the variables are normally distributed. Spearman correlation coefficient spans between -1 and +1 with 0 indicating no correlation. Correlation coefficients of -1 or +1 imply perfect monotonic relationship. Using the spearman function from the sub-package scipy.stats (13) the correlations between each of the two target genes (knockout genes) and the rest of the genes were calculated. The correlated genes were chosen for further analysis, all of which with FDR-adjusted p-values < 0.01 and |Spearman's ρ| > 0.4.

### Principal Component Analysis (PCA)

PCA as a dimensionality reduction technique was used to compare the gene expression profile dispersion of the bacteria based on the variations of their genes' expression levels. "pca" function from mixOmics R package was used for this purpose (14).

### Pathway analysis

BioCyc (15) database of microbial genomes and metabolic pathways was used to find the pathways each of the correlated genes are assigned to.

# Results

## Experimental results

Growth measures of bacterial cells with or without plasmid on LB and M9+Glycerol are presented in Fig. 1 and 2, respectively. Samples were taken within 24 hours at different time intervals. On LB media the preferential growth dynamics of knockout strains with or without *Ts*FDH over control strains were recognizable more clearly after eight hours post inoculation up to 24 hours. In samples grown on M9+Glycerol, we observed a pronounced shift in the growth divergence of FDH knockout strains with or without *Ts*FDH from control strains during earlier time intervals up to 24 hours compared with LB medium. On M9+Glycerol, BL21 with *Ts*FDH showed higher growth up to eight hours compared with the standard BL21. Moreover, on M9+Glycerol, both knockouts with *Ts*FDH showed relatively higer growth rates from 12 hours onward compared with the respective growth rates on LB. SDS-PAGE analysis of the strains confirms the expression of *Ts*FDH under the experimental conditions (Fig. S3).

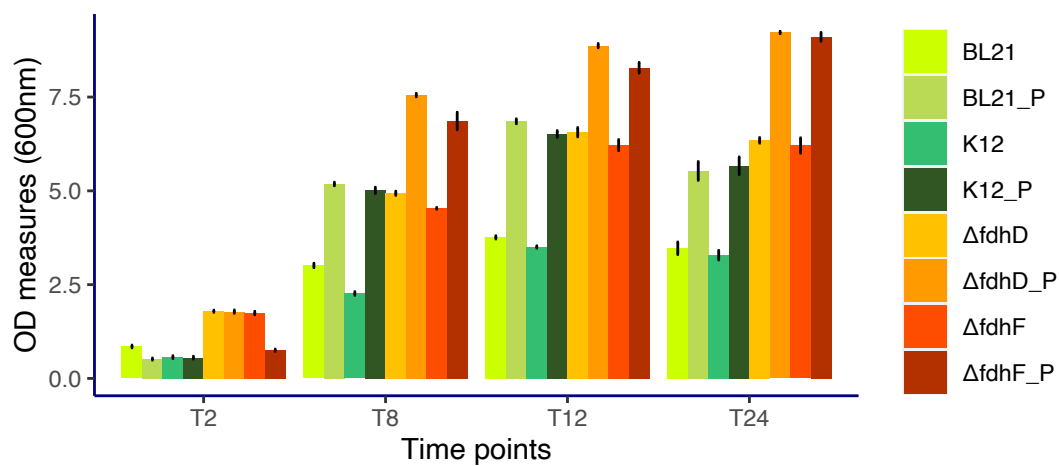

*Figure 1. Growth rate comparison among bacteria on LB medium at OD 600 nm at different time points.* At earlier time points there is no considerable difference between the bacterial species with and without TsFDH. From eight hours onward, there is a significant difference between all bacterial species with and without TsFDH.

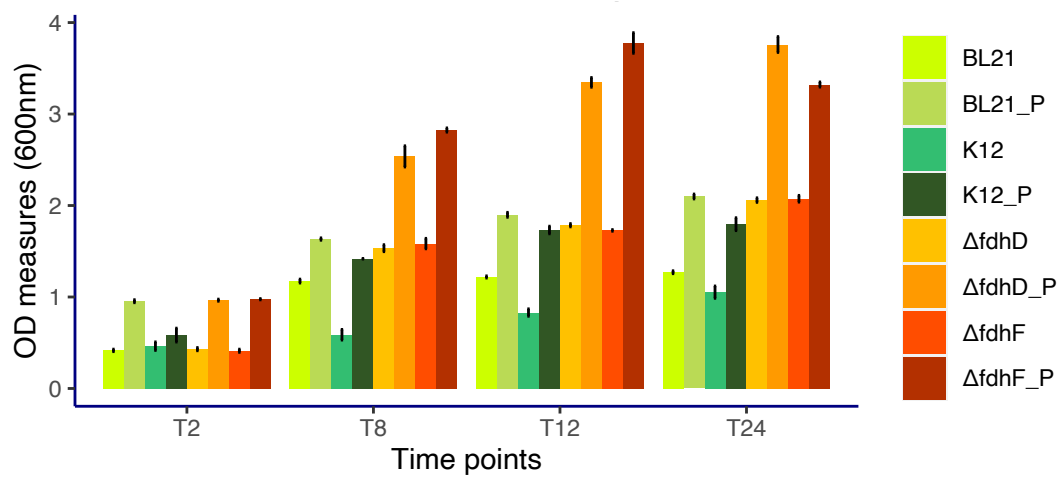

*Figure 2. Growth rate comparison among bacteria on M9+Glycerol medium at OD 600 nm at different time points. At earlier hours, unlike on LB medium, on M9+Glycerol the growth rates of the bacterial species with TsFDH are vividly higher than those without TsFDH except for K12. Particularly from 12 hours onward, the difference between the knockouts with and without TsFDH is relatively more accentuated compared with the growth rates on LB at the same time points.*

## In silico analysis results

### correlation analysis:

All correlations with each of the knockout genes were calculated (Table S3 and S4) and the ones with FDR-adjusted p-values < 0.01 were chosen (Table S5 and S6).

### PCA analysis:

According to the PCA results (Fig. S4a), it is postulated that the top anti-correlated genes for both knockouts are closely associated with one another. As a comparison, the PCA plot was also generated using all the genes, indicating that the other genes show more expression divergence. Moreover, the second PCA plot (Fig. S4b) could be an indication that the dataset, being based on different growth media, reflects a wide range of expression levels for different genes in each data point, which is critical for reflecting the correlations between the fluctuating gene expression levels.

## Metabolic pathways

Since the knockout of FDH main subunits is synonymous to absolute down regulation of the FDH gene in knockout strains, we initially focused on the significantly negatively correlated genes, which may reveal the genes that undergo up regulation accordingly. Using BioCyc database of microbial genomes and metabolic pathways, all the significantly negatively correlated genes (with FDR adjusted p-values < 0.01) were shown to be involved in tRNA charging pathway and tRNA processing pathway (PWY0-1479). To evaluate whether the mentioned pathways, are more frequent among the negatively or positively correlated gene, all the positively correlated genes assigned to the same pathways, i.e., tRNA charging and tRNA processing pathways, were selected for comparison. All the pathways other than the two mentioned pathways for both negatively and positively correlated genes were categorized as background pathways and were named "All other pathways". For better visualization of the distribution of all positive and negative correlation coefficients for both mentioned pathways the corresponding density plots were used (Fig 3. and 4.). The comparisons of the percentages of the number of correlated genes in each pathway category are given for both fdhD and fdhF genes (Table 2).

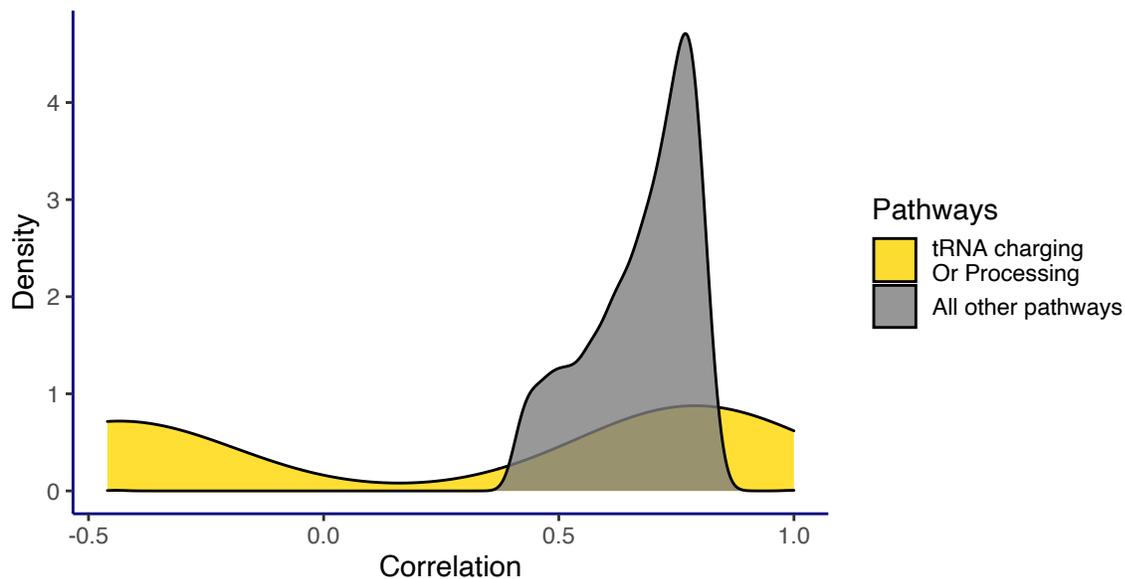

*Figure 3. Correlations distribution for fdhF.* *The distribution of all genes' correlations with fdhF for each pathway. The gold density plot represents the probability density of achieving either tRNA charging or tRNA processing for the corresponding correlation coefficients on the x axis. The gray density plot represents the probability density of getting any pathway other than the two mentioned ones for the corresponding correlation coefficients on the x axis.*

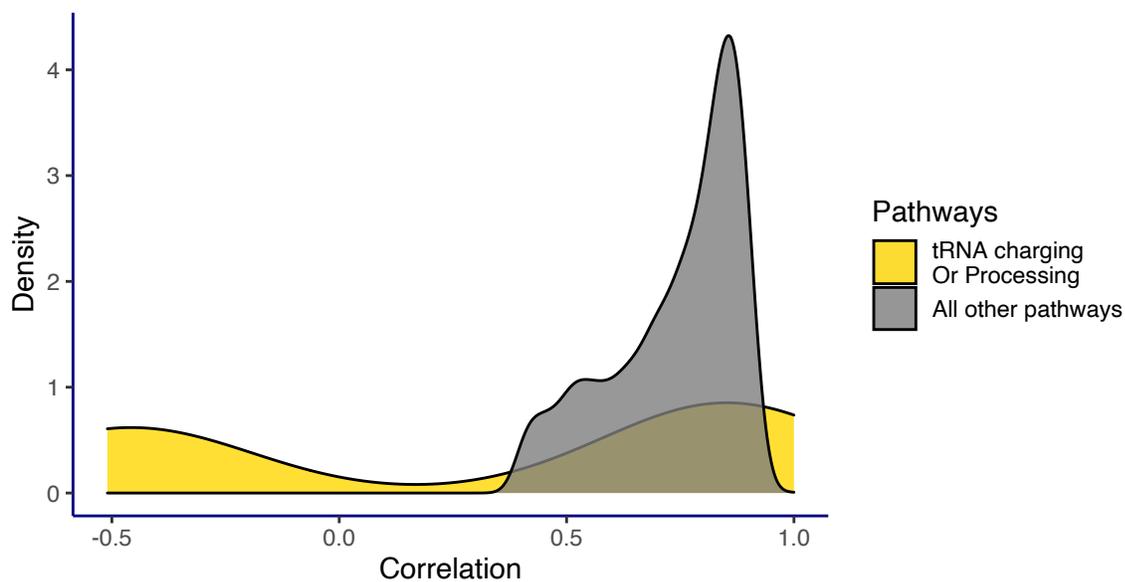

***Figure 4. Correlations distribution for fdhD.*** *The distribution of all genes' correlations with fdhD for each pathway. The gold density plot represents the probability density of achieving either tRNA charging or tRNA processing for the corresponding correlation coefficients on the x axis. The gray density plot indicates the probability density of getting any pathway other than the two mentioned ones for the corresponding correlation coefficients on the x axis.*

*Table 2. Correlations with fdhF and fdhD.* Comparison of the number of positively and negatively correlated genes assigned to the unique set of pathways.

| Knockout gene | Pathway | Negatively correlated genes | Positively correlated genes |
|---|---|---|---|
| fdhF | t-RNA charging or processing | 96% | 1% |
|  | All other pathways | 4% | 99% |
| fdhD | t-RNA charging or processing | 100% | 0.8% |
|  | All other pathways | 0% | 99.2% |

# Discussions

Increasing the growth efficacy of industrially important microorganisms is a novel goal in biotechnological applications. One of the strategies to boost the bacterial growth rate is to reduce the organic carbon leak, i.e., the release of $CO_2$ as one of the main end products in the respiration process. Different *E. coli*, strains are the workhorse for the production of some well-known biopharmaceuticals, like G-CSF, Romiplostim and Asparaginase. Therefore, *E. coli* is a suitable model microorganism and developing a strain of *E. coli* with higher growth rate is in demand particularly in Biotechnology.

Since the Kelvin cycle is the mainstream of $CO_2$ fixation pathway in plants, algae and cyanobacteria, most engineering efforts are directed towards the Kelvin cycle for converting $CO_2$ into valuable materials. Heterotrophic microorganisms generally do not assimilate $CO_2$ through the central metabolism (16). Over the past decade, there has been great success in the production of $CO_2$ derivatives, which have the potential to be used as fuel and valuable chemicals by bacteria (17). Previously, other researchers have approached this challenge by defining fermentation conditions and controlling aeration rates (18) or with genetically overexpressing ArcA transcription factor (19).

Here, we introduce a new approach by targeting one of the main enzymes responsible for converting organic formate into inorganic wasteful $CO_2$, i.e., FDH. There are three known FDHs in *E. coli*, namely, respiratory FDH, anaerobically expressed FDH and newly identified pressure induced FDH (FHL). FDHF is the cytosolic form, while FDHN and FDHO are membrane bound, with FDHN responsible for nitrogen cycle and FDHO active in sulfur metabolism (20). All these enzymes prefer the oxidation of formate into $CO_2$ under physiological conditions. Scanning BRENDA for FDHs with tendency towards the production of formate from $CO_2$ revealed that there are few candidate FDHs with relatively higher formate production ($CO_2$ reduction) catalytic efficiency. A comprehensive search of *E. coli's* metabolism using the Regulon DB (21) showed that the formate dehydrogenase enzymes were successfully expressed in recombinant form in *E. coli* (22). According to a study (16), *E. coli*'s FDHs have a strong tendency for regenerating $CO_2$ from formate. Among the studied formate dehydrogenases, *Ts*FDH has potential advantages as a biocatalyst in the field of $CO_2$ reduction (7). We hypothesized that perhaps the *E. coli* strains harboring *Ts*FDH could lead to developing bacterial strains for biotechnological applications with higher biomass to carbon source ratio owing to the possible lowering of carbon dioxide leakage. This possible solution has so far been remained out of sight and to the best of our knowledge not tried yet.

We expressed the recombinant fdhD gene from *Thiobacillus* in *Escherichia coli* strain K12 and BL21. In order to better compare the role of *Ts*FDH in growth efficacy on similar growth conditions (media), we also transformed two FDH knockout strains from K12 and BL21 with pET+TsFDH. Significant growth rate differences were observed between the knockout strains and the recombinant knockouts containing *Ts*FDH on either growth medium. On the other hand, both knockouts with *Ts*FDH showed vividly higher growth rate starting from 8 hours all the way to 24 hours on both media post induction. Additionally, the growth advantage of the control strains (K12 and BL21) with *Ts*FDH lasted up to 12 hours on both media. A possible reason for the less increase in growth rate of BL21 cells transformed with *Ts*FDH plasmid compared with the transformed knockouts is the presence of the original FDH in these strains. BL21 also contains IPTG as a gene expression inducer thanks to having a chromosomally encoded bacteriophage T7 RNA polymerase (T7 RNAP), which may be responsible for some metabolic leakage leading to less growth compared with K12-derived knockouts and transformed knockouts (23). The original K12 cells and mutants derived from this cell lack the complementary T7 RNA-polymerase.

In this study, we showed that removing either subunit of the wild type formate dehydrogenase gene from *E. coli* and its replacement with the formate dehydrogenase gene from *Thiobacillus* Sp. KNK65MA can increase the growth rate of *E. coli* cells. In a previous study by Palsson *et al.*, the whole-cell *in silico* model of *E. coli* metabolic network predicted that glycerol should be a preferred carbon source over glucose. However, the experimental findings were not consistent with the mentioned predictions. They indicated the adaptive evolution phenomenon for the bacteria to go from sub-optimal to the predicted optimal growth rate on glycerol (24–26). Concerning our results, the replacement of the native FDH with TsFDH might lead to potential metabolic rewirings leading to an increased glycerol efficiency as a carbon source. Our outcomes may suggest an initial clue to start more mechanistic metabolic investigations such as flux balance analysis and $CO_2$ leakage measurements to address these discrepancies between the *in silico* predictions and the experimental outcomes. Considering our *in silico* outcomes, we hypothesised that the omission of the original *E. coli* FDH may ultimately lead to the increased expression of some genes playing roles in tRNA charging and processing pathways. These *in silico* findings could be considered as a gene expression-level study related to our experimental observations; nonetheless, more detailed investigations are necessary to verify the hypotheses arising from our results.

# Conclusions

The main problems with studying most FDHs published so far are protein instability, sensitivity to oxygen and the low conversion rate. The FDH of this study (*Ts*FDH) shows some biochemical advantages over the previously studied FDHs such as Candida bolidini's FDH including higher turnover number and insensitivity to the environmental oxygen (17,27,28).

We showed that both knockouts and transformation with *Ts*FDH leads to an overall increased growth in all strains. In the initial incubation periods, the growth rates were approximately equal between transformed and non-transformed knockouts on LB medium, while in 8-24h the growth rates of the transformed and knockout bacteria were much higher than those of controls in both media, which implies the negative effect of wild type FDH gene on the *E. coli* growth rate. In other words, there would be a higher growth rate by eliminating the wild type FDH chains.

The increased growth rate of the transformed knockouts on both media might be a consequence of decreased $CO_2$ leakage due to less formate oxidation by *Ts*FDH compared with the wild type FDH. One plausible hypothesis regarding the relatively more accentuated growth of the transformed knockouts on glycerol medium could be the preferred utilization of glycerol after knockout and transformation besides the probable decrease of $CO_2$ leakage due to the above-mentioned conjecture. To evaluate these hypotheses, further studies are necessary such as $CO_2$ absorption/emission measurements and flux balance analysis (FBA). These observations could also be a starting point for more sophisticated molecular-level studies on *Ts*FDH contribution to the growth efficiency of *E. coli* on glycerol as the carbon source.

# List of abbreviations

PCA: Principal Component Analysis; FDH: Formate Dehydrogenase; TsFDH: Thiobacillus sp. KNK65MA; DE3: Escherichia coli BL21; CPM: Count Per Million

Acknowledgements
We acknowledge the fund by EU-ITN project ProteinFactory (MSCA-ITN-2014-ETN-642836) and the Swedish Research Council (Grant 2016-03798). We thank Payam Emami, Rui Benfeitas and Paulo Czarnewski at the National Bioinformatics Infrastructure Sweden (NBIS) at SciLifeLab for their fruitful discussions and help with the RNA-seq data analyses. We also thank Roghaieh Ghaderi Ternik for her help.

Authors' contributions
RR and BB conducted experiments. AAS and MHM helped with the design of experiments. SHA conceived and executed the bioinformatics analysis sections. AE helped with the bioinformatics analyses and visualizations. BB, RR and SHA wrote the manuscript. All authors read and approved the manuscript.

Funding
- International Cooperation for Applied Research Development (ICARD) grant to BB.

Availability of data and materials
- All the data supporting the conclusions of this paper are included in the context of the paper and the additional files.

# Declarations

Ethics approval and consent to participate
- Not applicable.

Consent for publication
- Not applicable.

Competing interests
-Not applicable.